\documentclass[onecolumn,aps,prd,groupedaddress]
              {revtex4}%
\usepackage{graphicx}
\usepackage{amsmath}
\usepackage{bm}
\usepackage{relsize}
\RequirePackage{xspace}

\begin{document}


\title{\mbox{}\\[10pt]
Radiative decays of charmonium into light mesons}

\author{Ying-Jia Gao, Yu-Jie Zhang, Kuang-Ta Chao}
\affiliation{ {\footnotesize Department of Physics, Peking
University,
 Beijing 100871}}




\begin{abstract}
We apply perturbative QCD to the radiative decays of charmonia
$J/\psi$ and $\chi_{cJ}$ into light mesons.  We perform a complete
numerical calculation for the quark-gluon loop diagrams involved
in these processes. The calculated $J/\psi$ decay branching ratios
into P-wave mesons $f_2(1270)$ and $f_1(1285)$ fit the data well,
while that of $f_0(980)$ (if treated as an $s\bar s$ meson) is
predicted to be $1.6\times 10^{-4}$, which implies that
$f_0(1710)$ can not be the $s\bar s$ or $(u\bar u+d\bar
d)/\sqrt{2}$ meson. Decays of P-wave charmonia $\chi_{cJ}\to
\rho(\omega, \phi)\gamma$ (J=0,1,2) are also studied, and the
branching ratio of $\chi_{c1}\to \rho\gamma$ is predicted to be
$1.4\times 10^{-5}$, which may be tested by CLEO-c and BESIII with
future experiments.
\end{abstract}


\maketitle


Decays of heavy quarkonium into light hadrons are forbidden
processes by the OZI rules. The radiative decays of $J/\psi$ into
light hadrons such as $J/\psi\to\gamma f_2(1270)$ are expected to
proceed via two virtual gluons that subsequently convert to light
mesons, with the photon emitted by the charm quarks (see Fig.1 for
the three Feynman diagrams). These processes are interesting since
they can provide useful information for understanding the
fundamental theory of strong interactions, namely, quantum
chromodynamics (QCD).

 \begin{figure*}[t]
\includegraphics[width=13cm]{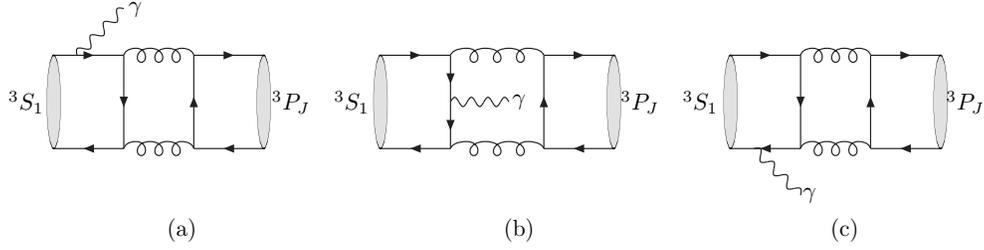}
\caption{Feynman diagrams for charmonium $(^3S_1)$ radiative
decays into light mesons $(^3P_J)$} \label{fvs}
\end{figure*}

In this letter, we will report calculations for radiative decays
of charmonia $J/\psi$ and $\chi_{cJ}$ into various light mesons
based on nonrelativistic quantum chromodynamics. Hopefully, the
$J/\psi$ mass is large enough for using perturbative QCD to
calculate these processes.

Differing from previous works (see,
e.g.,\cite{kor,Krammer:1978qp}), we will perform a complete
numerical calculation for the quark-gluon loop diagrams involved
in these processes.

Here, it is emphasized that in our approach the description of
$J/\psi$ as a nonrelativistic $c\bar c$ bound state is a good
approximation; while treating the light mesons such as $f_2(1270)$
as nonrelativistic $q\bar q$ bound states $(q=u, d, s)$ with
constituent quark masses $m_u=m_d\approx 350~MeV,~~m_s\approx
500~MeV$ is only an approximation for describing the
nonperturbative dynamics that has been used in the nonrelativistic
constituent quark models rather successfully, and was also used in
previous studies for these radiative decay
processes\cite{kor,Krammer:1978qp}. In fact, if the current quark
masses for the light quarks, which will vanish in the chiral
limit, were used in the quark-gluon loop diagram, we would
encounter difficulties with soft and collinear divergences.


In the following, we will first focus on the radiative decays of
$J/\psi$ to P-wave light mesons such as $f_2(1270)$, $f_1(1285)$,
and $f_0(980)$. We adopt the descriptions of nonrelativistic bound
states, i.e., both $J/\psi$ and light mesons are described by the
color-singlet non-relativistic wave functions. As widely accepted
assignments we assume that $f_2(1270)$ and $f_1(1285)$ are mainly
composed of $(u\bar u+d\bar d)/\sqrt{2}$ (neglecting the mixing
with $s\bar s$ for simplicity). But for $f_0(980)$, there are many
possible assignments such as the tetraquark state, the $K\bar K$
molecule, and the P-wave $s\bar s$ dominated state (for related
discussions on $f_0(980)$ and other scalar mesons, see, e.g., the
topical review--note on scalar mesons in \cite{Pdg} and
\cite{close}). Since experimental data show that $D_s^{+}\to
f_0\pi^{+}$ has a large branching ratio (BR)\cite{Pdg}, we prefer
to assign $f_0(980)$ as an $s\bar s$ dominated P-wave state as a
tentative choice.

In the calculations, the spin projection operators for
$^{3}S_{1}$, $^{1}S_{0}$ and $^{3}P_{J}$ states and their
relations can be found in \cite{liu,Kuhn}.
We use  {\tt FeynCalc} \cite{FeynCalc} for the tensor reduction
and {\tt LoopTools}\cite{LoopTools} for the numerical evaluation
of infrared safe integrals. We follow the way in
~Ref.\cite{fourrank} to deal with five-point functions and high
tensor loop integrals that can not be calculated by {\tt
LoopTools} and {\tt FeynCalc}, such as
 \begin{eqnarray}
\!\!\!\!\!E_{\alpha\beta\rho\sigma}&=&\int\mathrm{{d}}^Dk
\frac{k_\alpha k_\beta k_\rho k_\sigma}{k^2[(k+p_q)^2-m_q^2](k+2
p_q)^2 [(p_c+k)^2-m_c^2][(k+2 p_q-p_c)^2-m_c^2]},
\end{eqnarray}
where $p_c$ is the momentum of $c$ quark, $p_q$  the momentum of
light quark, $m_c$ the charm quark mass, and $m_q$ the constituent
light quark mass of $f_J$ meson. For the constituent quark masses,
we choose $m_u=m_d=0.35$ GeV, $m_s=0.50$ Gev,
$m_c=1.5\,\mbox{GeV}$.

Using the theoretical expression for the widths of
$f_2\to\gamma\gamma$,

 \begin{eqnarray}
\Gamma_{f_2(1270)\to \gamma\gamma}^{(th)}&=&\frac{6 N_c}{5}
(Q_u^2+Q_d^2)^2 \alpha^2 \frac{|\mathcal{R}'_P(0)|^2}{m^4},
\hspace{1cm} \Gamma_{f_2'(1525)\to \gamma\gamma}^{(th)}=\frac{12
N_c}{5} Q_s^4 \alpha^2 \frac{|\mathcal{R}'_P(0)|^2}{m^4},
\end{eqnarray}
where $N_c$=3 is the color number, $\alpha$=$\frac{1}{137}$, and
fitting them with their experimental values 2.6~KeV and 0.081~KeV
for $f_2(1270)$ and $f_2'(1525)$ respectively\cite{Pdg},
we get
\begin{equation}
|\mathcal{R}_{f_2(1270)}'(0)|^2 =6.6\times 10^{-4}\,\,\,\,
\mbox{GeV$^5$},\hspace{1cm} |\mathcal{R}_{f_2'(1525)}'(0)|^2
=1.1\times 10^{-3}\,\,\,\, \mbox{GeV$^5$}.
\end{equation}
For charmonium, the value of the wave function at the origin is
taken to be \cite{Quig}:
 \begin{eqnarray}
|\mathcal{R}_{J/\psi}(0)|^2=0.81~GeV^3, \hspace{1cm}
|\mathcal{R}^{\prime}_{\chi_{cJ}}(0)|^2=0.075~GeV^5.
\end{eqnarray}

%



In the calculation, the meson masses are taken to be $M=2m_q$ in
the nonrelativistic limit. For the $J/\psi$ decay we take
$\alpha_s(2m_c)=0.26$. Our results are listed in the Table I.

\begin{table}[tb]
\begin {center}
\begin{tabular}{ |c|c|c|c|c|c|}
 \hline
process &$J/\psi\to\gamma f_0(980)$&$J/\psi\to\gamma
f_1(1285)$&$J/\psi\to\gamma f_2(1270)$&$J/\psi\to\gamma
f_1'(1420)$&$J/\psi\to\gamma f_2'(1525)$
\\\hline
$BR_{th}$&$1.6\times10^{-4}$&$7.0\times10^{-4}$&$8.7\times10^{-4}$&$1.8\times10^{-4}$&$2.0\times10^{-4}$\\
$BR_{ex}$&$\backslash$&$(6.1\pm0.8)\times10^{-4}$&$(13.8\pm1.4)\times10^{-4}$
&$(7.9\pm1.3)\times10^{-4}$&$(4.5^{+0.7}_{-0.4})\times10^{-4}$\\
\hline
\end{tabular}
\caption{Numerical results for ${\cal B}(J/\psi\to f_J\gamma)$.}
 \label{table1}
\end {center}
\vspace{-0.5cm}
\end{table}

%

Then we will give the contributions of different helicities. As in
Ref.\cite{JPMa}, we choose the moving direction of $f_J$ as the
z-axis, and introduce three polarization vectors:
 \begin{eqnarray}
\omega^{\mu}(1)&=&\frac{-1}{\sqrt{2}}(0,1,i,0),\hspace{1cm}
\omega^{\mu}(-1)\!\!=\!\!\frac{1}{\sqrt{2}}(0,1,-i,0),\hspace{1cm}
\omega^{\mu}(0)=\frac{1}{m}(|\textbf{k}|,0,0,k^0), \end{eqnarray}
thus we can characterize the polarization tensor $\epsilon^{\mu
\nu}(\lambda)$ of $f_2$. With the same parameters, we give the
branching ratios for different helicity states in Table II. The
corresponding values of helicity parameters $x$ and $y$ are
$x=0.46$, $y=0.23$. Recently, new experimental data for the
contributions of different helicities in process $J/\psi\to\gamma
f_2(1270)$ have been given by the BES Collaboration
\cite{besjpsi}: $x=0.89 \pm 0.02 \pm 0.10$ and $y=0.46 \pm 0.02
\pm 0.17$ (see also\cite{Pdg}). It is about 2 times larger than
our results. But if we use a larger constituent quark mass, e.g.
$m_u=M(1270)/2$, we will get substantially increased values
$x=0.79$ and $y=0.58$ (also see Ref.\cite{kor}). We emphasize that
the helicity parameters are very sensitive to the light quark
masses, and hence very useful in clarifying the decay mechanisms.
Note that if $m_u/m_c\to 0$, we will have $x\to 0$ and $y\to 0$,
but this is inconsistent with data.

\begin{table}[tb]
\begin {center}
\begin{tabular}{ |c|c|c|c|c|c|}
 \hline
process&$|\lambda|$&$BR$&process&$|\lambda|$&$BR$\\\hline
$J/\psi\to\gamma f_1(1285)$&$1$&$0.18\times 10^{-4}$&$J/\psi\to\gamma f_1'(1420)$&$1$&$0.12\times 10^{-4} $\\
&$0$&$6.81\times 10^{-4}$&&$0$&$1.64\times10^{-4}$\\\hline
$J/\psi\to\gamma f_2(1270)$&$2$&$0.38\times 10^{-4}$&$J/\psi\to\gamma f_2'(1525)$&$2$&$0.21\times 10^{-4}$\\
&$1$&$1.44\times 10^{-4}$&&$1$&$0.51\times 10^{-4}$\\
 &$0$&$6.91\times 10^{-4}$&&$0$&$1.27\times 10^{-4}$
\\\hline
\end{tabular}
\caption{Results for $J/\psi\to f_J\gamma$ with different helicity
states. }
 \label{table1}
\end {center}
\vspace{-0.5cm}
\end{table}


From Table I we see BR of $J/\psi\to f_0(980)\gamma$ (assuming
$f_0(980)$ to be $s\bar s$) is predicted to be $1.6\times10^{-4}$.
Actually, this calculation can also apply for $f_0(1710)$ (if
treated as $s\bar s$) with an even smaller BR due to the phase
space correction. However, experimentally  BR of $J/\psi\to
f_0(1710)\gamma$ is larger than $9\times 10^{-4}$\cite{Pdg}. This
implies that the $f_0(1710)$ can not be a pure $s\bar s$ or
$(u\bar u+d\bar d)/\sqrt{2}$ scalar meson (for the latter the BR
is predicted to be about $4\times 10^{-4}$). One possible
explanation is that the $f_0(1710)$ has a substantial gluonic
component due to $q\bar q$-glueball mixing.

Next, we discuss $J/\psi$ radiative decays to $\eta$ and $\eta'$.
Assume $\eta'$ and $\eta$  can be written as admixtures of
flavor-singlet $\psi_1$ and octet $\psi_8$ :

 \begin{eqnarray}
\psi_1&=&\frac{1}{\sqrt{3}}(u\bar u + d\bar d + s\bar s),\,\,\,
\psi_8=\frac{1}{\sqrt{6}}(u\bar u + d\bar d - 2 s\bar s),\\
\eta&=&\psi_8 \cos{\theta} -\psi_1 \sin{\theta},\,\,\, \eta'=\psi_8
\sin{\theta} +\psi_1 \cos{\theta}, \end{eqnarray}
where $\theta$ is the mixing angel, taken to be a widely accepted
value $\theta=-20^{\circ}$\cite{gasser}. For the $\eta(\eta')$,
the wave function at the origin may be determined from the
leptonic decay of vector mesons $V\to e^{+}e^{-}$~
($V=\phi,\rho$), with the assumption that in the nonrelativistic
quark model the wave functions of the $0^{-+}$ and $1^{--}$ mesons
should be equal in the nonrelativistic limit. Using
\begin{eqnarray} \Gamma(\phi(1020)\to e^{+}e^{-})&=&N_c
Q_s^2 \alpha^2 \frac{|\mathcal{R}(0)|^2}{3 m_s^2}
= (1.27\pm 0.04)~\mbox{KeV},
\end{eqnarray} we can get
$|\mathcal{R}(0)|^2$=0.054 $\mbox{GeV$^3$}$ for $s\bar s$.
Similarly we can get  $|\mathcal{R}(0)|^2$=0.032 \mbox{GeV$^3$}
for $u\bar u$ and $d\bar d$ from $\rho$ leptonic decay. The decay
branching ratios of $J/\psi\to\gamma\eta(\eta')$ are then
calculated and listed in Table\ref{etaetaprime}.

\begin{table}[tb]
\begin {center}
\begin{tabular}{|c|c|c|}
 \hline
process&$BR_{th}$&$BR_{ex}$\\ \hline
$J/\psi\to\gamma\eta$&$3.0\times10^{-4}$&$(8.6\pm0.8)\times10^{-4}$\\
$J/\psi\to\gamma\eta'$&$2.3\times10^{-3}$&$(4.31\pm0.3)\times10^{-3}$\\
\hline
\end{tabular}
\caption{Branching ratios for $J/\psi\to \eta(\eta')\gamma$.}
 \label{etaetaprime}
\end {center}
\vspace{-0.5cm}
\end{table}
We can see that for $J/\psi\to\gamma\eta(\eta')$ the simple
nonrelativistic quark model calculations fit the data rather well,
and the values are compatible with other approaches e.g. the
$\eta(\eta')-\eta_c$ mixing mechanism based on the QCD axial
anomaly\cite{chao}.

Another interesting processes are decays of P-wave charmonia
$\chi_{cJ}\to \rho(\omega, \phi)\gamma$ (J=0,1,2), where the
photon is emitted by the light quarks but not the charm quarks, in
contrast to the above discussed $J/\psi$ radiative decays.  The
calculated branching ratios are shown in Table~\ref{table1}. In
particular, the branching ratio of $\chi_{c1}\to \rho\gamma$ is
predicted to be $1.4\times 10^{-5}$, which may be tested by CLEO-c
and BESIII in future experiments.
\begin{table}[tb]
\begin {center}
\begin{tabular}{ |c|c|c|c|c|c|}
 \hline
process&$BR_{th}$&process&$BR_{th}$&process&$BR_{th}$
\\\hline
 $\chi_{c2}\to\gamma\rho$&$4.4\times 10^{-6}$&$\chi_{c2}\to\gamma\omega$&$5.0\times 10^{-7}$
 &$\chi_{c2}\to\gamma\phi$&$1.1\times 10^{-6}$\\
 $\chi_{c1}\to\gamma\rho$&$1.4\times 10^{-5}$&$\chi_{c1}\to\gamma\omega$&$1.6\times 10^{-6}$
 &$\chi_{c1}\to\gamma\phi$&$3.6\times 10^{-6}$\\
 $\chi_{c0}\to\gamma\rho$&$1.2\times 10^{-6}$&$\chi_{c0}\to\gamma\omega$&$1.3\times 10^{-7}$
 &$\chi_{c0}\to\gamma\phi$&$4.6\times 10^{-7}$\\
\hline
\end{tabular}
\caption{Predicted branching ratios for $\chi_{cJ}\to\gamma
\rho(\omega,\phi)$.}
 \label{table1}
\end {center}
\vspace{-0.5cm}
\end{table}

In this letter, we apply perturbative QCD to the radiative decays
of $J/\psi$ and $\chi_{cJ}$ into light mesons by treating the
light mesons such as $f_2(1270)$ as nonrelativistic $q\bar q$
states as in the nonrelativistic constituent quark models. We
perform a complete numerical calculation for the quark-gluon loop
diagrams, where the light quark masses are taken to be the
constituent quark masses. The calculated $J/\psi$ decay branching
ratios to the P-wave mesons $f_2(1270)$ and $f_1(1285)$ fit the
data well (but the observed helicity parameter $x$ for $f_2(1270)$
favors even larger light quark masses). Moreover, the results for
S-wave mesons $\eta(\eta')$ seem to be also rather good. Our
calculation does not support $f_0(1710)$ being the $s\bar s$ or
$(u\bar u+d\bar d)/\sqrt{2}$ meson, while the possible $s\bar s$
interpretation for $f_0(980)$ can be tested. The predicted
branching ratios for $\chi_{cJ}\to \rho(\omega, \phi)\gamma$
(J=0,1,2) are generally small, and may be tested by CLEO-c and
BESIII with future experiments.

However, we must bear in mind that for $J/\psi$ decays the
applicability of perturbative QCD is marginal, and the
nonperturbative effects should be important. Moreover, the
successes of nonrelativistic constituent quark models for light
hadrons have not been fully understood from QCD. The constituent
quark model description for $\eta$ and $\eta'$ is even more
questionable. In this paper, to evaluate the quark-gluon loop
diagrams and to avoid the soft and collinear divergences caused by
massless light quarks, we have tried to use the constituent quark
masses as in the nonrelativistic quark model for light mesons.
Hopefully, the good (if not accidental) fit of the calculation
presented here with the experimental data can shed light on a
deeper understanding of $J/\psi$ decays in QCD, which evidently
deserve further studies.

We thank Ce Meng for valuable discussions. This work was supported
in part by the National Natural Science Foundation of China (No
10421503), and the Key Grant Project of Chinese Ministry of
Education (No 305001).


\end{document}